\documentclass[english,a4paper]{article}
\usepackage{dcolumn}
\usepackage{bm}
\usepackage{graphicx}
\usepackage{amsmath}
\usepackage{amsfonts}
\usepackage{amssymb}
\usepackage{amsthm}
\usepackage{amstext}
\usepackage{amsbsy}
\usepackage{amsopn}
\usepackage{amscd}
\usepackage{amsxtra}

\def\openone{\leavevmode\hbox{\small1\kern-3.3pt\normalsize1}}

\DeclareMathOperator*{\cn}{cn}
\DeclareMathOperator*{\sn}{sn}
\DeclareMathOperator*{\dn}{dn}
\usepackage[usenames,dvipsnames]{color}

\begin{document}
\title{Classical and quantum rotation numbers of asymmetric top molecules}
\author{K. Hamraoui, L. Van Damme, P. Marde\v si\'c\footnote{Institut de Math\'ematiques de Bourgogne, UMR 5584 CNRS-Universit\'e de Bourgogne Franche-Comt\'e, 9 Av. A. Savary, BP 47870 21078 Dijon Cedex, France}, D. Sugny\footnote{Laboratoire Interdisciplinaire Carnot de
Bourgogne (ICB), UMR 6303 CNRS-Universit\'e Bourgogne-Franche Comt\'e, 9 Av. A.
Savary, BP 47 870, F-21078 Dijon Cedex, France and Institute for Advanced Study, Technische Universit\"at M\"unchen, Lichtenbergstrasse 2 a, D-85748 Garching, Germany, dominique.sugny@u-bourgogne.fr}}

\maketitle

\begin{abstract}
We study the classical and quantum rotation numbers of the free rotation of asymmetric top molecules. We show numerically that the quantum rotation number converges to its classical analog in the semi-classical limit. Different asymmetric molecules such as the water molecule are taken as illustrative example. A simple approximation of the classical rotation number is derived in a neighborhood of the separatrix connecting the two unstable fixed points of the system. Furthermore, a signature of the classical tennis racket effect in the spectrum of asymmetric molecules is identified.
\end{abstract}

\section{Introduction}
Different theoretical and computational works have clearly shown the benefit of classical analyses for revealing and understanding the properties of quantum molecular spectra~\cite{childbook,efstathiou,gutzwiller,maslov,ishikawa,cuisset2012,cuisset2013,sugnyhcn,sugnyreview,harter1984,miller1978}. A well-known example is given by Hamiltonian monodromy, which is the simplest topological obstruction to the existence of global action angle variables in classical integrable systems \cite{cushman,Duistermaat,dullin2008,dullin2009}. Monodromy has become a standard and useful tool to describe the global properties of molecular spectra \cite{co2mono,child2007,efstathiou2004,arango2005,kozin2003}. A non-trivial quantum monodromy prevents the existence of global good quantum numbers to describe the spectrum of quantum systems. The connection between classical and quantum monodromies has been rigorously established in the semi-classical regime~\cite{ngoc1999}. Recently, different kinds of generalized monodromy, such as fractional monodromy \cite{efstathiou,nekhoroshev2006,sugny2008} and bidromy \cite{sadovskii2007,assemat2010,efstathiou2010}, have been defined and their presence shown in model and molecular systems. Free rotation of asymmetric molecules has no monodromy~\cite{cushman}, even if non trivial dynamics occurs in presence of external electric fields~\cite{arango2008}. In the same direction, it has recently been shown that the non rigidity of molecular tops can be at the origin of a bifurcation which leads to the destabilisation of one of the two stable axes of inertia~\cite{cuisset2012,cuisset2013}. This change of stability has profound implications in the structure of the corresponding rotational spectrum. Finally, note that different analogies have been recently established between the rotation of a rigid body and the dynamics of quantum systems~\cite{SR,opatrny}.

\begin{figure}[!htpb]
\centering
    \includegraphics[width=\linewidth]{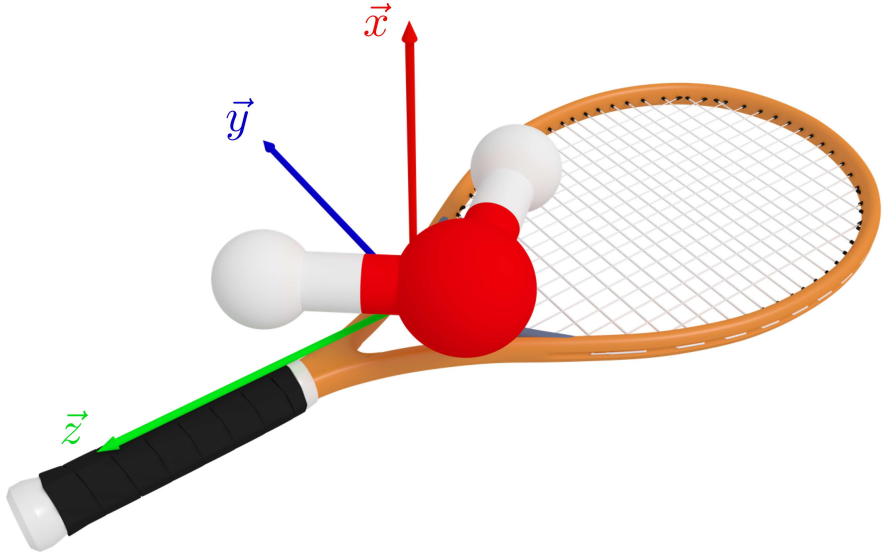}
\caption{(Color online) Schematic picture of the classical-quantum correspondence between the water molecule and a tennis racket. The body-fixed frame $(\vec{x},\vec{y},\vec{z})$ is defined for a tennis racket and for the water molecule.}
\label{figracket}
\end{figure}

A geometric description of the rotational spectrum of asymmetric top molecules can be obtained using the rotation number~\cite{cushman,efstathiou}. The classical rotation number is a dynamical object of a classical integrable Hamiltonian system which describes locally the twist of a Hamiltonian trajectory lying on a torus. This abstract mathematical concept has a simple interpretation in the rotation of a tennis racket, which is a specific asymmetric rigid body~\cite{arnold,goldstein,landau}. A schematic representation is given in Fig.~\ref{figracket}. As an illustrative example, we consider the Tennis Racket Effect (TRE), a classical geometric phenomenon occurring in the free rotation of a rigid body~\cite{cushman,vandamme2015,MSA91,petrov}. More precisely, TRE describes what happens when a tennis racket is tossed into the air while imparting a rotation about an axis. A tennis racket has three inertia axes as schematically represented in Fig.~\ref{figracket}. The axis $\vec{z}$ is along the handle of the racket, $\vec{y}$ lies in the plane of the head of the racket and is orthogonal to $\vec{z}$, while $\vec{x}$ is orthogonal to the head of the racket. A moment of inertia is associated with each axis with the convention $I_z< I_y< I_x$. The axes $\vec{z}$ and $\vec{x}$ are stable so nothing unexpected happens for a rotation about theses axes. If we now spin the racket about its transverse axis, a surprising effect is observed. In addition to the intended $2\pi$- rotation about its transverse axis, the racket will almost always perform an unexpected $\pi$- flip about its handle. In other words, when the racket is caught, the initial bottom side will be facing up. The unstable character of the intermediate axis is responsible for this effect~\cite{cushman,vandamme2015,MSA91,petrov}. The racket exactly goes back to its initial position after 2 TREs. During this motion, the handle has made a rotation of $4\pi$. This variation corresponds to the classical rotation number for a trajectory satisfying the TRE and is therefore a characterization of this geometric effect. The rotation number has a quantum analog which has been rigorously defined in~\cite{san2017}. This number can be computed directly from the spectrum of the quantum system.

This work is aimed at studying the classical and quantum rotation numbers of asymmetric top molecules. The semi-classical regime is used to establish the link between the two rotation numbers. An approximate expression of the classical rotational number is derived close to the separatrix connecting the two unstable fixed points. The quantum rotation number is defined from a set of good quantum numbers labelling the spectrum. However, two physical choices of good quantum numbers can be made. The so-called oblate and prolate quantum numbers correspond respectively to the limit of purely rotating and oscillating motions~\cite{zare}. We investigate the relation between the two rotation numbers and we show that their difference is equal to $1$ for any asymmetric molecule.  At this point, an intriguing question is to identify a signature of TRE in the spectrum of the quantum system. This classical-quantum correspondence can be uncovered in the semi-classical limit~\cite{gutzwiller,childbook,child2007}. As could be expected, the quantum rotation number of asymmetric top molecules is close to 2 for energy levels in the neighborhood of the separatrix where the TRE is classically observed (note that, by definition, the classical and the quantum rotation numbers differ by a factor $2\pi$). The approximate expression of the classical rotation number is used to estimate the distance to the separatrix where the TRE occurs.

The paper is organized as follows. In Sec.~\ref{sec2}, we recall the equations governing the free rotation of a rigid body. Section~\ref{sec3} is dedicated to the definition of the classical and quantum rotation numbers in the case of asymmetric molecules. In Sec.~\ref{sec4}, we present the computation of the quantum rotational spectrum, as well as the semi-classical limit. In particular, we show numerically that the quantum rotation number converges to its classical analog in the semi-classical regime. Some conclusions and discussions are presented in Sec.~\ref{sec5}. Technical computations about the classical dynamics of a rigid body are reported in Appendices~\ref{appb}, \ref{appa} and \ref{appc}.
\section{Free rotation of a rigid body}\label{sec2}
\begin{figure}[!htpb]
\centering
      \includegraphics[width=\linewidth]{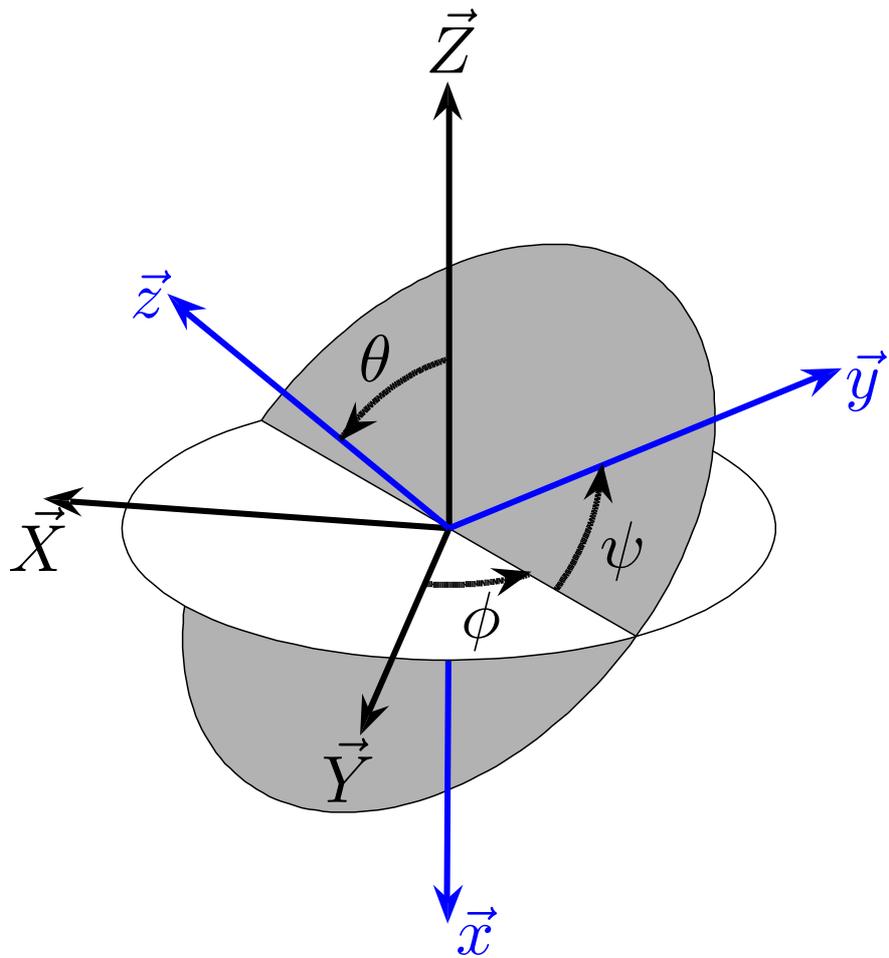}
\caption{(Color online) Definition of the Euler angles used to describe the position of the body-fixed frame $(\vec{x},\vec{y},\vec{z})$ in the laboratory frame $(\vec{X},\vec{Y},\vec{Z})$ (see the text for details).}
\label{figangle}
\end{figure}

\begin{figure}[!htpb]
\centering
      \includegraphics[width=0.8\linewidth]{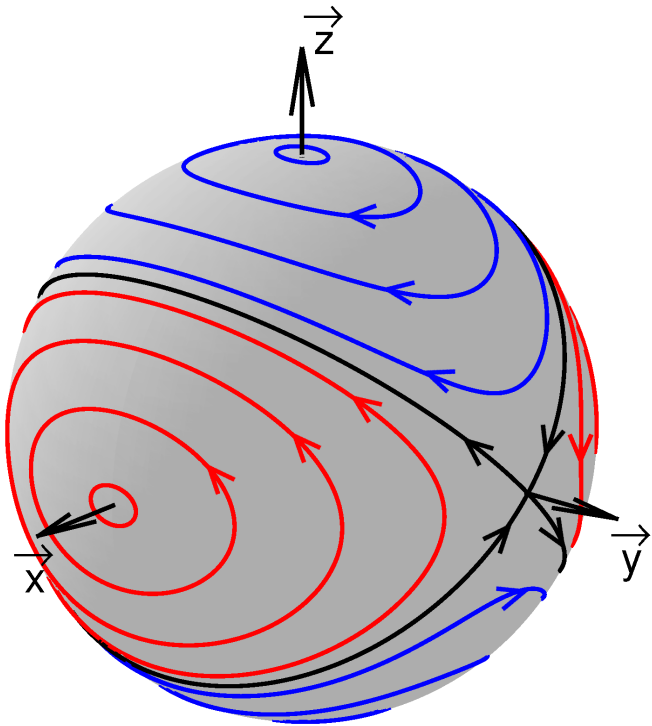}
\caption{(Color online) Dynamics of the angular momentum
of a rigid body in the body-fixed frame $(\vec{x},\vec{y},\vec{z})$. The red (dark gray) and blue (light gray)
lines represent respectively the rotating and oscillating trajectories of the angular momentum. The solid black line is the separatrix.}
\label{figportrait}
\end{figure}

The free rotation of a rigid body can be described through a particular set of Euler angles~\cite{landau,goldstein,vandamme2015} defined in Fig.~\ref{figangle}, which displays the position of the body-fixed frame $(\vec{x},\vec{y},\vec{z})$ with respect to the laboratory frame $(\vec{X},\vec{Y},\vec{Z})$.
In this system of coordinates, the precession of the handle of the racket about the angular momentum $\vec{J}$ along the $\vec{Z}$- axis is given by the angle $\phi$, while the flip of the head of the racket is measured by $\psi$. In a TRE experiment, it can be assumed that $\theta$ is of the order of $\pi/2$. Along a trajectory such that $\Delta\phi=2\pi$, the TRE manifests by a variation $\Delta\psi=\pi$~\cite{cushman,vandamme2015}. The components of $\vec{J}$ can be expressed in the body-fixed frame as $J_x=-J\sin\theta\cos\psi$, $J_y=J\sin\theta\sin\psi$ and $J_z=J\cos\theta$, where $J=|\vec{J}|$~\cite{landau}. The dynamics of the angular momentum is governed by the Euler equations:
\begin{equation}\label{eqeuler}
\dot{\vec{J}}=\vec{J}\times \vec{\Omega},
\end{equation}
with $\Omega_x=J_x/I_x$, $\Omega_y=J_y/I_y$ and $\Omega_z=J_z/I_z$. In order to map the quantum results into the classical framework, we use the convention $A=\frac{1}{2I_x}$, $B=\frac{1}{2I_y}$ and $C=\frac{1}{2I_z}$, with $A<B<C$, where $A$, $B$ and $C$ are the rotational constants of the molecule. The solutions of the differential system \eqref{eqeuler} are recalled in Appendix~\ref{appb}. The different trajectories
that can be followed by $\vec{J}$ are displayed in Fig.~\ref{figportrait}. The classical phase space has a simple structure made of a
separatrix which is the boundary between two families of trajectories, the rotating and the oscillating ones. Each family of trajectories is
distributed around a stable fixed point~\cite{landau,goldstein}.
The dynamics of the Euler angles are governed by~\cite{arnold}:
\begin{equation}
\begin{cases}
\dot{\theta}=2J(B-A)\sin\theta\sin\psi\cos\psi ,\\
\dot{\phi}=2J(B\sin^2\psi+A\cos^2\psi),\\
\dot{\psi}=2J(C-B\sin^2\psi-A\cos^2\psi)\cos\theta.
\end{cases}
\label{eqeulerangles}
\end{equation}
A Hamiltonian description of the dynamics can be derived from the classical Hamiltonian $H=A J_x^2+B J_y^2+C J_z^2$~\cite{childbook} from which we deduce the expressions of the momenta (see Appendix~\ref{appa}):
\begin{equation}
\begin{cases}
p_\psi=J_z \\
p_\theta=J_x\sin\psi+J_y\cos\psi \\
p_\phi=-J_x\sin\theta\cos\psi+J_y\sin\theta\sin\psi+J_z\cos\theta ,
\end{cases}
\label{eqmomenta}
\end{equation}
with the relation:
\begin{equation}\label{eqj}
J^2=p_\theta^2+\frac{1}{\sin^2\theta}(p_\phi-p_\psi\cos\theta)^2+p_\psi^2.
\end{equation}
Using Eq.~\eqref{eqj}, we introduce a canonical transformation from the set of variables $(p_\theta,\theta,p_\psi,\psi,p_\phi,\phi)$ to $(J,\alpha_J,K,\alpha_K, M,\alpha_M)$ defined by the generating function $S(\theta,\psi,\phi,J,K,M)$~\cite{childbook}, which satisfies:
$$
p_\theta=\frac{\partial S}{\partial \theta},~p_\phi=\frac{\partial S}{\partial \phi},~p_\psi=\frac{\partial S}{\partial \psi}=K.
$$
Straightforward computations show that $H$ can be expressed as follows:
\begin{equation}
H=(J^2-K^2)(A\cos^2\alpha_K+B\sin^2\alpha_k)+CK^2.
\label{eqH}
\end{equation}
This classical system is Liouville-integrable~\cite{arnold} since it has as many constants of the motion as the number of degree of freedom, that is 3: $H$, $J$ and $M$. Note that $H$ does not depend on $M$, so that only two degrees of freedom can be considered. As displayed in Fig.~\ref{figem}, the energy-momentum diagram (EM) is a useful way to visualize the global dynamics of an integrable system. It corresponds to all the possible values of $H=E$ and $J$.
The position of the stable fixed points of the free rotation of a rigid body, $E=AJ^2$ and $E=CJ^2$, delimits the boundary of the EM and of the accessible phase space, while the separatrix, the trajectory connecting the two unstable fixed points, is defined by $E=BJ^2$~\cite{goldstein}.
The separatrix distinguishes the two families of trajectories, namely the rotating and the oscillating ones for $E>BJ^2$ and $E<BJ^2$, respectively.
\section{Classical and quantum rotation numbers}\label{sec3}
\begin{figure}[!htpb]
\centering
      \includegraphics[width=\linewidth]{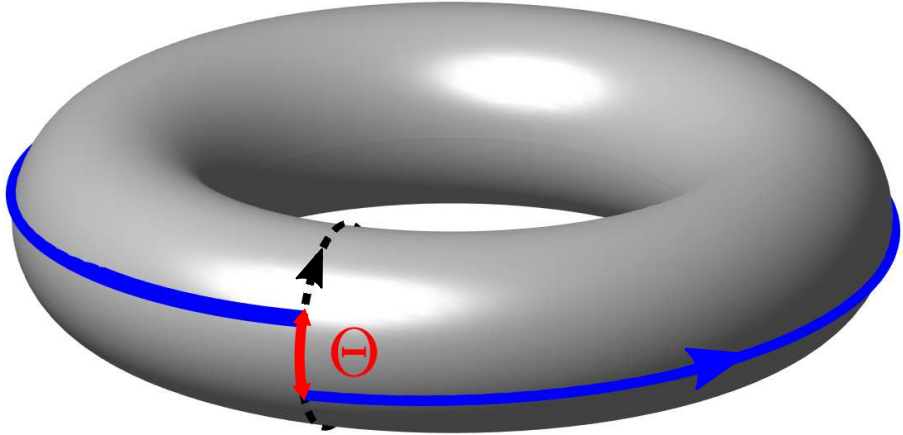}
\caption{(Color online) Classical definition of the rotation number. The blue (dashed gray) solid and black dashed lines represent respectively the flow of $H$ and $J$.}
\label{figrn}
\end{figure}

According to the Liouville-Arnold theorem~\cite{arnold}, any regular point of the EM is associated with a torus of the phase space. We consider a point $I$ of this torus and the orbit of this point under the action of $J$. This trajectory is topologically equivalent to a circle since $J$ is an action coordinate of the system. The flow of $H$ starting from $I$ gives another trajectory lying on the torus which is not periodic and intersects the orbit of $I$ for the first time $T$ in $F$. The classical rotation number $\Theta_{cl}$ is defined as the variation of the angle $\alpha_J$ along the orbit: $\Theta_{cl}=\alpha_J(F)-\alpha_J(I)$. A schematic representation of this construction is given in Fig.~\ref{figrn}. The rotation number allows us to define explicitly the second action coordinate $\mathcal{I}$ of the system as $d\mathcal{I}=-\frac{\Theta_{cl}}{2\pi}dJ+\frac{T}{2\pi}dH$~\cite{efstathiou,cushman}.
In the semi-classical limit, the quantum lattice of the EM can be described locally by an elementary cell defined by:
\begin{equation}
\begin{cases}
\Delta J=\Delta J \\
\Delta \mathcal{I}=-\frac{\Theta_{cl}}{2\pi} \Delta J+\frac{T}{2\pi} \Delta H,
\end{cases}
\end{equation}
where $J$ is the first action coordinate. In the quantum case, we denote by $E_{j,p}$ the energies of the quantum Hamiltonian, where $(j,p)$ are the two good quantum numbers labelling the spectrum, $j$ and $p$ being associated respectively with $J$ and $\mathcal{I}$. Note that the same value of $p$ is assigned to the quasi-degenerate energy levels.
Along a line with the same value of $p$ for two consecutive levels in $j$, we have $\Delta J=1$ and $\Delta \mathcal{I}=0$. We deduce that:
$$
\Delta E=E_{j+1,p}-E_{j,p}\simeq\frac{\Theta_{cl}}{T},
$$
where the equality holds true in the semi-classical limit. For two consecutive levels in $p$ with the same $j$, we obtain $\Delta J=0$ and $\Delta \mathcal{I}=1$, i.e.:
$$
\Delta E=E_{j,p+1}-E_{j,p}\simeq \frac{2\pi}{T}.
$$
We finally arrive at the definition of the quantum rotation number:
$$
\Theta_Q=\frac{E_{j+1,p}-E_{j,p}}{E_{j,p+1}-E_{j,p}},
$$
which can be directly computed from the Hamiltonian spectrum. Note that the mathematical definition of the quantum rotation number has been introduced in~\cite{san2017}. In the semi-classical regime, we get: $\Theta_Q=\frac{\Theta_{cl}}{2\pi}+O(h)$. The action coordinates are not uniquely defined~\cite{arnold}. For a second action of the form $\mathcal{I}'=\mathcal{I}+nJ$, $n\in\mathbb{Z}$, the corresponding rotation numbers can be expressed as $\Theta_{cl}'=\Theta_{cl}-2n\pi$ and $\Theta_Q'=\Theta_Q-n$.

\begin{figure}[h!]
\centering
\includegraphics[scale=0.5]{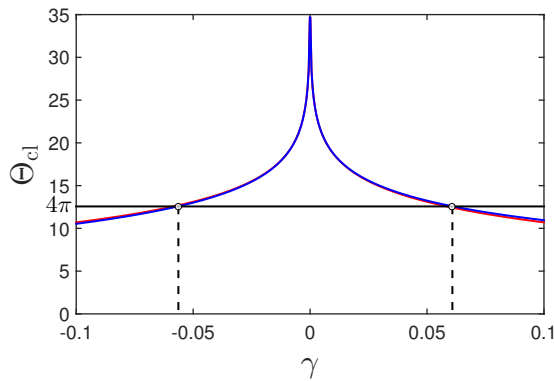}
\caption{(Color online) Evolution for the water molecule of the classical rotation number $\Theta_{cl}$ (black or blue line) and of its approximation (gray or red line) given by Eq.~\eqref{apprtheta} as a function of $\gamma$. The horizontal solid line corresponds to the TRE for which $\Theta_{cl}=4\pi$.\label{figappr}}
\end{figure}

For a rigid body, the first return time $T$ is the period of the motion of the angular momentum. By definition, the rotation number $\Theta_{cl}$ corresponds to the variation of the angle $\alpha_J$ during the time $T$, but, as shown in Appendix~\ref{appa}~\cite{bates2005}, $\Theta_{cl}$ is also given by the variation of $\phi$ along the Hamiltonian flow. Furthermore, a geometric interpretation of the classical rotation number $\Theta_{cl}$ can be obtained by using the Montgomery phase~\cite{montgomery,natario}. This latter is a geometric phase associated with the angular momentum of a rigid body, which is analog to the Berry phase in quantum physics~\cite{berryphase,nakahara}. For sake of completeness, a complete derivation is given in Appendix~\ref{appc}. The rotation number can be expressed as follows:
\begin{equation}\label{thetaphase}
\Theta_{cl}=\frac{2ET}{J}-\mathcal{A},
\end{equation}
where $\frac{2ET}{J}$ and $\mathcal{A}$ are respectively the dynamical and geometric contributions to $\Theta_{cl}$. As defined in Appendix~\ref{appc}, $\mathcal{A}$ is the solid angle swept out by the angular momentum during a loop. Starting from Eq.~\eqref{thetaphase}, an approximate expression of $\Theta_{cl}$ can be derived in a neighborhood of the separatrix (See Appendix~\ref{appc} for details):
\begin{equation}\label{apprtheta}
\Theta_{cl}\simeq \alpha-\beta \ln (|\gamma |),
\end{equation}
where $\alpha$ and $\beta$ are two functions of the rotational constants $A$, $B$ and $C$, and $\gamma$,
the distance to the separatrix, $E=BJ^2(1+\gamma)$. The accuracy of this approximation is shown in Fig.~\ref{figappr} for $|\gamma|\leq 0.1$. Note the logarithmic divergence of $\Theta_{cl}$ on the separatrix. The asymptotic expression of $\Theta_{cl}$ is the same in the oscillating ($\gamma<0$) and rotating ($\gamma>0$) areas.
For the rotation of a tennis racket, we know that a racket exactly goes back to its initial position after two TREs, that is for the angular variations $\Delta\psi = 2\pi$ and $\Delta \phi = 4\pi$. For trajectories in a neighborhood of the separatrix for which the TRE is observed, we deduce therefore that $\Theta_{cl}=4\pi$ and that $\Theta_Q$ goes to 2 when $h\to 0$. Equation~\eqref{apprtheta} allows to estimate the distance to the separatrix needed to get a $4\pi$- variation. Figure~\ref{figappr} shows that the TRE can be observed for $|\gamma|\simeq 0.05$.
\section{Quantum rotational spectrum}\label{sec4}
\begin{figure}[!htpb]
\centering
      \includegraphics[width=\linewidth]{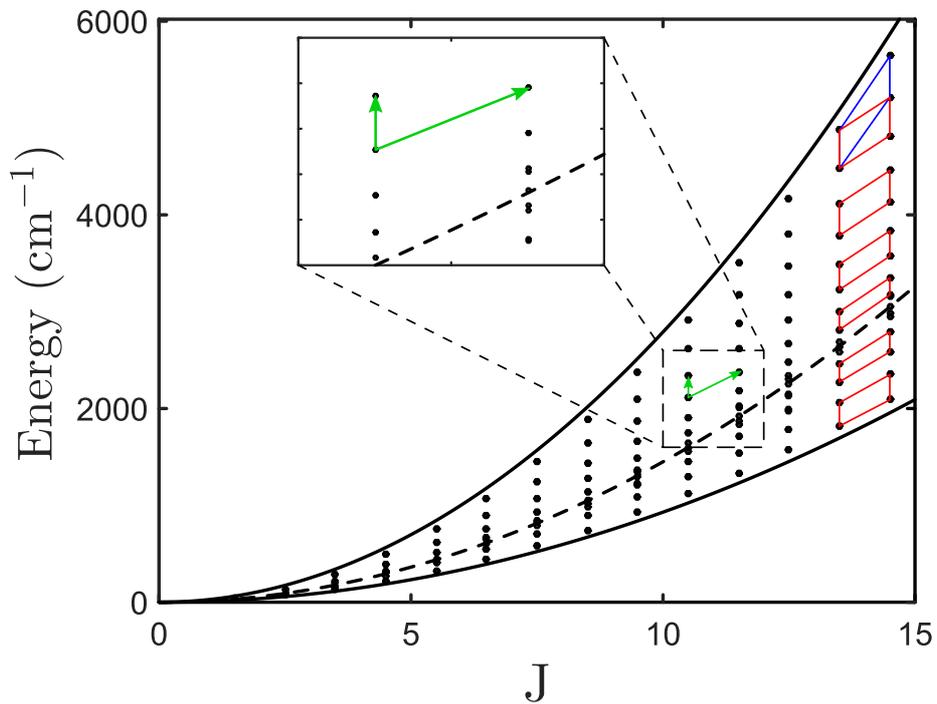}
\caption{(Color online) Quantum energy momentum diagram of the water molecule for $h=1$. The small insert shows the two arrows used to compute the quantum rotation number $\Theta_Q$. For this cell, we have $\Theta_Q\simeq 2.45$. The dashed line depicts the position of the separatrix and the solid lines the boundary of the accessible EM. The blue (black) and red (dark gray) cells represent respectively the rotating and oscillating definitions of the second good quantum number $p$.}
\label{figem}
\end{figure}

This paragraph is aimed at computing the quantum rotation number from the rotational spectrum of the molecule. The quantum dynamics is governed by the Hamiltonian $\hat{H}=h^2(A\hat{J}_x^2+B\hat{J}_y^2+C\hat{J}_z^2)$~\cite{zare,landauQ} (the \emph{hat} symbol is used to distinguish the quantum operators from the classical ones). The constant $h$ is an effective dimensionless Planck constant which can, at least theoretically, be modified at will to increase the density of energy levels and to reach the semi-classical regime. The value $h=1$ corresponds to the physical problem. Units are chosen so that $\hbar =1$. For the water molecule, numerical values are taken to be $A=9.3$, $B=14.5$ and $C=27.9$ in cm$^{-1}$ \cite{landolt}. We also introduce the components $(\hat{J}_X,\hat{J}_Y,\hat{J}_Z)$ of the angular momentum in the laboratory frame. In the Wigner basis $|j,k,m\rangle$, with $j\geq 0$, $-j\leq k\leq j$, $-j\leq m\leq j$, the angular momentum satisfies:
\begin{equation}
\begin{cases}
\hat{J}_z|j,k,m\rangle = k|j,k,m\rangle \\
\hat{J}_Z|j,k,m\rangle = m|j,k,m\rangle \\
\hat{J}^2|j,k,m\rangle = j(j+1)|j,k,m\rangle .
\end{cases}
\end{equation}
In the $\{|j,k,m\rangle\}$ basis, the Hamiltonian matrix has a tri-diagonal structure in which $j$ and $m$ are good quantum numbers~\cite{zare}. Note that this spectrum which is computed numerically does not depend on the value of $m$. The quantization rule for the angular momentum is $J=h(j+\frac{1}{2})$~\cite{childbook}. As displayed in Fig.~\ref{figem}, the spectrum of the system can be represented as a lattice of points in a two-dimensional space in terms of $J$ and the energy $E$. However, this calculation is not sufficient to construct the quantum version of the rotation number. A choice has to be made to define the second quantum number $p$.

Two physical options correspond to the limit of a purely rotating or oscillating motion, where $E\simeq CJ^2$ or $E\simeq AJ^2$ respectively. In other words, the quantum levels of a given value of $j$ are labelled by the quantum number $p$ in increasing (resp. decreasing order) from $E=AJ^2$ (resp. $E=CJ^2$).

In molecular physics, this corresponds to the prolate and oblate quantum numbers~\cite{zare}. As can be seen in Fig.~\ref{figem}, two elementary cells can then be defined from these two definitions~\cite{ngoc1999,nekhoroshev2006}. Each cell is characterized by a quantum and a classical (in the semi-classical regime) rotation numbers, namely $\Theta^{(R)}$ and $\Theta^{(O)}$. A semiclassical analysis and the regular Bohr-Sommerfeld rules are needed to establish the exact correspondence between the quantum spectrum and the classical dynamics~\cite{landauQ,tabor}. Note that some energy levels too close to the separatrix cannot be described by these regular rules~\cite{colin}. For these levels, $\Theta_Q$ is not defined.

As in the monodromy phenomenon, we consider a vertical parallel transport of the oscillating cell~\cite{efstathiou2004,ngoc1999}. We assume that the cell can be transported through the separatrix, this generalized transport was defined for fractional monodromy in~\cite{nekhoroshev2006,sugny2008}. We show in Fig.~\ref{figem} that the rotating and the moved oscillating cells are different. The inconsistency between the two definitions comes from the fact that the two classical rotation numbers $\Theta_{cl}^{(R)}$ and $\Theta_{cl}^{(O)}$ are not equal in the limit $\gamma\to 0$, but satisfy the relation:
\begin{equation}\label{eqlim}
\lim_{\gamma\to 0}|\Theta_{cl}^{(R)}-\Theta_{cl}^{(O)}|=2\pi.
\end{equation}
This difference can be understood from the description of the rotation number in terms of the Montgomery phase. The geometric contribution to $\Theta_{cl}$ is the area $\mathcal{A}$ of the surface between the trajectory of the angular momentum and the equator (see Appendix~\ref{appc} for details).
In the limits $E\to AJ^2$ and $E\to CJ^2$, it is clear that $\mathcal{A}$ goes respectively to 0 and $2\pi$, showing therefore that Eq.~\eqref{thetaphase} corresponds to the oscillating definition (or to its continuous extension across the separatrix) of the classical rotation number.

In the rotation definition, $\mathcal{A}^{(R)}$ is the surface around the $z$- axis delimited by the trajectory of the angular momentum. We have $|\mathcal{A}^{(R)}-\mathcal{A}|=2\pi$. Since the dynamical phases are the same on the two sides of the separatrix for $|\gamma |\ll 1$, we finally deduce the relation \eqref{eqlim}, which is valid for any asymmetric molecule. 

\begin{figure}[!htpb]
\centering
      \includegraphics[width=\linewidth]{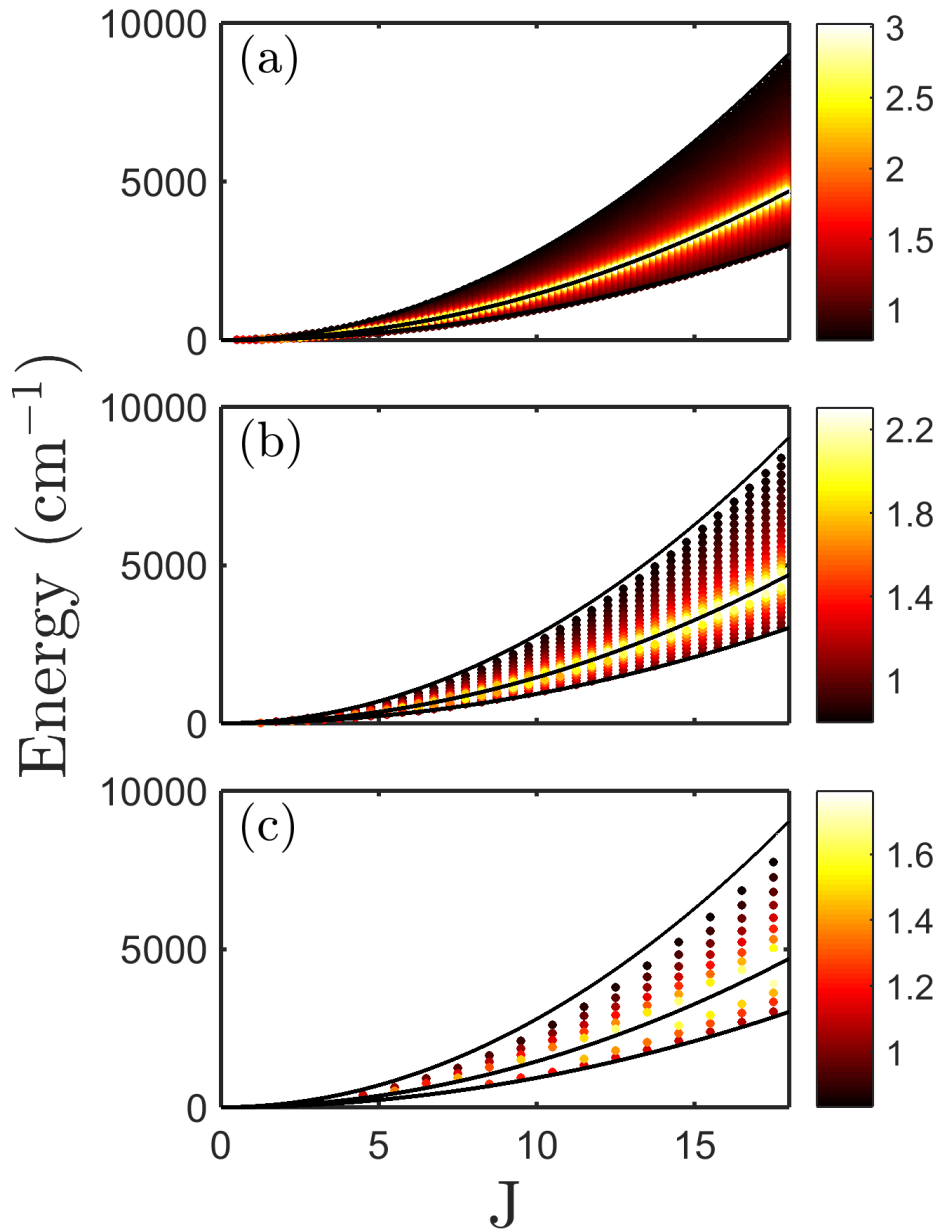}
\caption{(Color online) Quantum rotation number of the water molecule  as a function of $E$ and $J$ for $h=1$ (c), $h=0.5$ (b) and $h=0.2$ (a).}
\label{figapp}
\end{figure}

\begin{figure}[!htpb]
\centering
      \includegraphics[width=\linewidth]{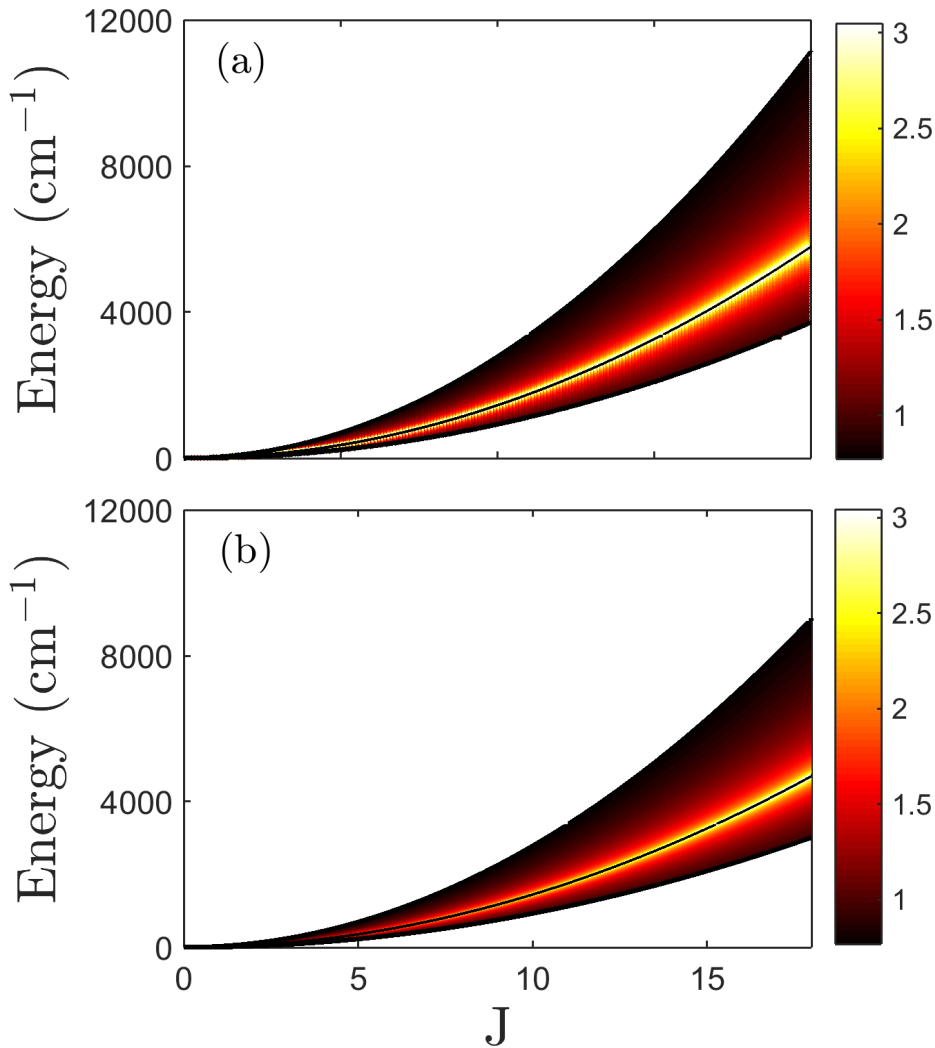}
\caption{(Color online) Quantum rotation number (a) for $h=0.1$ and classical rotation number divided by $2\pi$ (b) as a function of $E$ and $J$ for the water molecule.}
\label{figqrn}
\end{figure}

We now analyze the evolution of the quantum rotation number when the Planck constant $h$ goes to 0. Different spectra of the water molecule are displayed in Fig.~\ref{figapp} for different values of $h$. The connection between the quantum and the classical rotation numbers is illustrated in Fig.~\ref{figqrn}. We use here the oscillating definition for the quantum rotation number. As could be expected, a similar evolution is observed for the two quantities for small values of $h$. We observe in Fig.~\ref{figqrn} that $\Theta_Q$ converges to $\Theta_{cl}$ when $h$ goes to 0.

\begin{figure}[!htpb]
\centering
      \includegraphics[width=\linewidth]{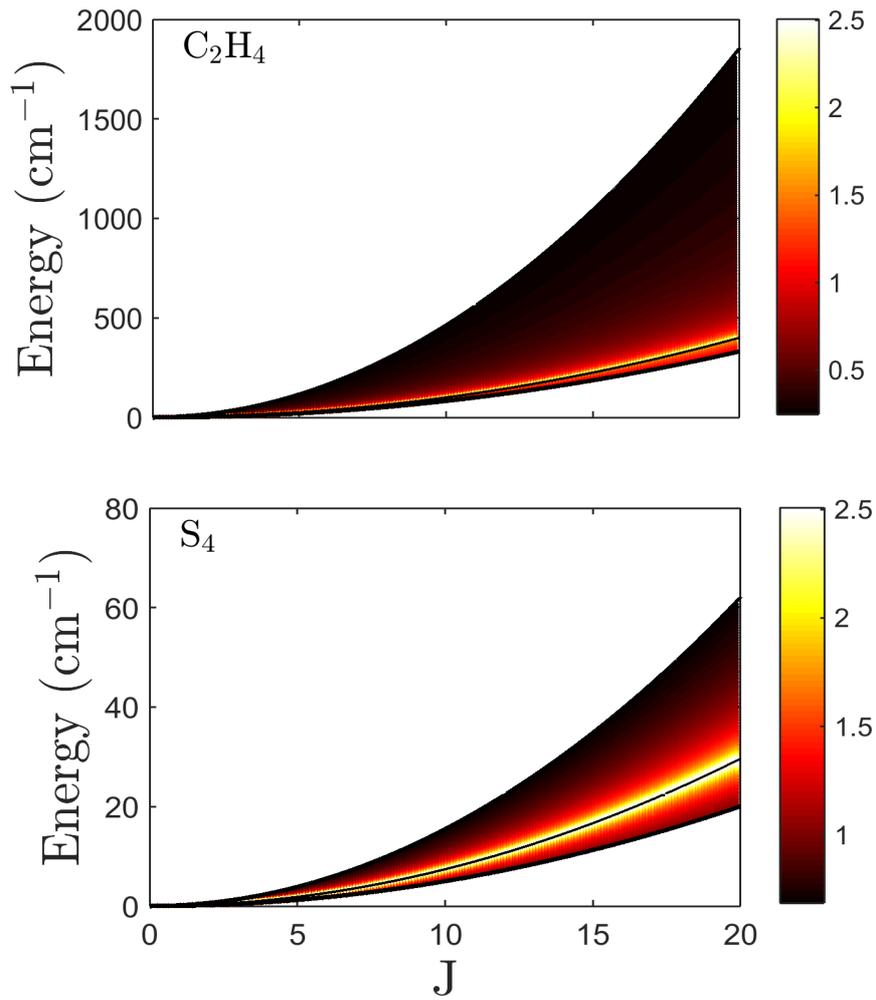}
\caption{(Color online) Quantum rotation number of the ethylene molecule (top) and of the $S_4$ molecule (bottom) for $h=0.1$ as a function of $E$ and $J$. Numerical values in cm$^{-1}$ are taken to be $A=0.828$, $B=1.001$ and $C=4.64$ for C$_2$H$_4$ and $A=0.0501$, $B=0.0741$ and $C=0.1553$ for S$_4$.}
\label{figother}
\end{figure}

For the water molecule, Fig.~\ref{figqrn} shows that $\Theta_Q$ is close to the value 2 in a neighborhood of the separatrix. This signature of TRE can also be exhibited in other asymmetric top molecules as shown in Fig.~\ref{figother}, highlighting the general character of this property. The distance to the separatrix for which $\Theta_{cl}=4\pi$ can be estimated with Eq.~\eqref{apprtheta}.

\section{Conclusion}\label{sec5}
We have studied in this work the classical and quantum rotation numbers in the free rotation of asymmetric top molecules. We have clarified the definition of the quantum rotation number based on the different choices of good quantum numbers labelling the rotational spectrum. A numerical analysis has revealed that the quantum rotation number converges to its classical analog in the semi-classical regime. We have also identified a signature of TRE in the rotational spectrum of such molecules. This effect manifests itself by an integer quantum rotation number in the neighborhood of the separatrix where the TRE can be observed. It can be exhibited in a larger number of molecules. At this point, it would be interesting to study other signatures of this classical property in the quantum dynamics, such as the evolution of a wavepacket starting initially in the neighborhood of the unstable equilibrium point of asymmetric top molecules. Coherent states could be used to be as close as possible as the classical trajectories~\cite{morales1999}. The final goal of this research project could be to propose an experimental demonstration of this effect by means of femtosecond laser fields~\cite{rouzee,lee,glaserreview}\\ \\
\textbf{Acknowledgment}\\
D. Sugny acknowledges support from the PICS program of the CNRS. The work of D. Sugny has been done with the support of the Technische Universit\"at M\"unchen – Institute for Advanced Study, funded by the German Excellence Initiative and the European Union Seventh Framework Programme under grant agreement 291763. We thank H.R. Jauslin, S.J. Glaser and S.V. Ngoc for helpful discussions.



\appendix


\section{Time evolution of the classical angular momentum}\label{appb}
We recall in this paragraph the time evolution of the angular momentum during a free rotation of a rigid body~\cite{landau,goldstein}. Starting from Eq.~\eqref{eqeuler}, it can be shown that:
\begin{equation}
\begin{cases}
\dot{J}_x=2J_yJ_z(C-B),\\
\dot{J}_y=-2J_xJ_z(C-A),\\
\dot{J}_z=2J_xJ_y(B-A).
\end{cases}
\end{equation}
The solutions of these equations are given in Tab.~\ref{tab1}.
\begin{table}[h!]
\begin{center}
\begin{tabular}{|c|c|}
\hline
Rotation & Oscillation \\
$E-BJ^2>0$ & $E-BJ^2<0$ \\
\hline
\hline
\small
$\begin{aligned}&J_x=-\sqrt{\frac{CJ^2-E}{C-A}}\cn(\chi,m)\\ &J_y=\sqrt{\frac{CJ^2-E}{C-B}}\sn(\chi,m) \\& J_z=\sqrt{\frac{E-AJ^2}{C-A}}\dn(\chi,m) \end{aligned}$ & \small
$\begin{aligned}&J_x=\sqrt{\frac{CJ^2-E}{C-A}}\dn(\chi,m)\\ &J_y=\sqrt{\frac{E-AJ^2}{B-A}}\sn(\chi,m) \\& J_z=\sqrt{\frac{E-AJ^2}{C-A}}\cn(\chi,m)\end{aligned}$\\
\hline
\hline
\small
$\begin{aligned}& \chi=\omega t+\rho \\ &\omega=2\sqrt{(C-B)(E-AJ^2)}\\ &m=\frac{(B-A)(CJ^2-E)}{(C-B)(E-AJ^2)} \end{aligned}$ &\small $\begin{aligned}& \chi=\omega t+\rho \\ &\omega=2\sqrt{(B-A)(CJ^2-E)}\\ &m=\frac{(C-B)(E-AJ^2)}{(B-A)(CJ^2-E)}\end{aligned}$\\
\hline
\end{tabular}
\end{center}
\caption{Analytic solutions of the Euler equations in the rotating and oscillating cases.\label{tab1}}
\end{table}
\section{Hamiltonian description of the Euler top}\label{appa}
We consider the standard definition of the Euler angles \cite{landau}. In the body-fixed frame, we have:
\begin{equation}
\begin{cases}
\Omega_x=-\dot{\phi}\sin\theta\cos\psi+\dot{\theta}\sin\psi \\
\Omega_y=\dot{\phi}\sin\theta\sin\psi+\dot{\theta}\cos\psi \\
\Omega_z=\dot{\phi}\cos\theta+\dot{\psi},
\end{cases}
\end{equation}
where the $\Omega_i$ are the angular velocities along the three directions of the frame. Note that the coordinates of the angular momentum are given by:
$$
J_x=I_x\Omega_x,~J_y=I_y\Omega_y,~J_z=I_z\Omega_z.
$$
The classical Hamiltonian of the system can be written as follows:
$$
H=\frac{1}{2}[I_x\Omega_x^2+I_y\Omega_y^2+I_z\Omega_z^2]=\frac{1}{2}[\frac{J_x^2}{I_x}+\frac{J_y^2}{I_y}+\frac{J_z^2}{I_z}].
$$
or
$$
H=AJ_x^2+BJ_y^2+CJ_z^2,
$$
in terms of the rotational constants. It is then straightforward to deduce the corresponding momenta:
\begin{equation}
\begin{cases}
p_\psi=\frac{\partial H}{\partial \dot{\psi}}=J_z \\
p_\theta=\frac{\partial H}{\partial \dot{\theta}}=J_x\sin\psi+J_y\cos\psi \\
p_\phi=\frac{\partial H}{\partial \dot{\phi}}=-J_x\sin\theta\cos\psi+J_y\sin\theta\sin\psi+J_z\cos\theta .
\end{cases}
\end{equation}
We obtain that:
\begin{equation}
\begin{cases}
J_x=\sin\psi p_\theta-\frac{\cos\psi}{\sin\theta}p_\phi+p_\psi\frac{\cos\psi}{\tan\theta} \\
J_y=\cos\psi p_\theta+\frac{\sin\psi}{\sin\theta}p_\phi-p_\psi\frac{\sin\psi}{\tan\theta} \\
J_z=p_\psi.
\end{cases}
\end{equation}
This leads to the relation:
\begin{equation}
J^2=p_\theta^2+\frac{1}{\sin^2\theta}(p_\phi-p_\psi\cos\theta)^2+p_\psi^2,
\label{eqj0}
\end{equation}
which can also be expressed as
\begin{equation}
J^2=p_\theta^2+\frac{1}{\sin^2\theta}(p_\phi^2-2p_\phi p_\psi\cos\theta+p_\psi^2).
\label{eqj2}
\end{equation}
We consider a canonical transformation from the set of variables $(p_\theta,\theta,p_\psi,\psi,p_\phi,\phi)$ to $(J,\alpha_J,K,\alpha_K, M,\alpha_M)$ defined by the generating function $S(\theta,\psi,\phi,J,K,M)$~\cite{childbook}. The function $S$ satisfies:
$$
p_\theta=\frac{\partial S}{\partial \theta},~p_\phi=\frac{\partial S}{\partial \phi},~p_\psi=\frac{\partial S}{\partial \psi}=K.
$$
Using Eq.~\eqref{eqj0}, we obtain:
$$
p_\theta=[J^2-\frac{1}{\sin^2\theta}(M^2-2KM\cos\theta+K^2)]^{1/2},
$$
and we deduce that $S$ can be written as:
$$
S=M\phi+K\psi+\int p_\theta d\theta .
$$
This leads to:
\begin{equation}
\begin{cases}
\alpha_J=\frac{\partial S}{\partial J}=\int\frac{\partial p_\theta}{\partial J}d\theta, \\
\alpha_K=\frac{\partial S}{\partial K}, \\
\alpha_M=\frac{\partial S}{\partial M},
\end{cases}
\end{equation}
which gives:
\begin{equation}
\begin{cases}
\alpha_J=\arccos [\frac{J^2\cos\theta -MK}{\sqrt{(J^2-K^2)(J^2-M^2)}}], \\
\alpha_K=\psi-\arccos [\frac{K\cos\theta-M}{\sin\theta\sqrt{J^2-K^2}}], \\
\alpha_M=\phi-\arccos [\frac{M\cos\theta-K}{\sin\theta\sqrt{J^2-K^2}}].
\end{cases}
\end{equation}
With the relations:
$$
\frac{K\cos\theta-M}{\sin\theta}=\sqrt{J^2-K^2}\cos(\alpha_K-\psi),
$$
and
\begin{eqnarray*}
& & \sqrt{J^2-\frac{1}{\sin^2\theta}(K\cos\theta-M)^2-K^2} \\
& & =-\sqrt{J^2-K^2}\sin(\alpha_K-\psi),
\end{eqnarray*}
we obtain:
\begin{equation}
\begin{cases}
J_x=\sqrt{J^2-K^2}\cos(\alpha_K),\\
J_y=-\sqrt{J^2-K^2}\sin(\alpha_K),\\
J_z=K.
\end{cases}
\end{equation}
The variables $(K,\alpha_K)$ describe the dynamics of the angular momentum $\vec{J}$ in the body-fixed frame.
With the convention:
\begin{equation}
\begin{cases}
J_x=-J\sin\theta\cos\psi,\\
J_y=J\sin\theta\sin\psi,\\
J_z=J\cos\theta,
\end{cases}
\end{equation}
we deduce the following identification:
$$
K=J\cos\theta;~\alpha_K=-\psi.
$$
The Hamiltonian $H$ can be expressed in terms of the new set of coordinates as follows:
$$
H=(J^2-K^2)(A\cos^2\alpha_K+B\sin^2\alpha_k)+CK^2.
$$
A first action of the system is given by $J$, the second $\mathcal{I}$ can be written as:
$$
\mathcal{I}=\int_\delta Kd\alpha_K,
$$
where $\delta$ is the loop projection of the Hamiltonian flow in the $(J_x,J_y,J_z)$- space.

The time evolution of the angle $\alpha_J$ can be obtained by using the relation $K=J\cos\theta$. We have:
$$
\cos(\alpha_J)=\frac{J-M}{\sqrt{J^2-M^2}}\cot(\theta),
$$
and
$$
\cos(\phi-\alpha_M)=\frac{M-J}{\sqrt{J^2-M^2}}\cot(\theta),
$$
and we deduce that $\alpha_J=\pi+\phi-\alpha_M$. Since
$$
\dot{\alpha}_M=\frac{\partial H}{\partial M}=0,
$$
we finally arrive at:
$$
\Theta_{cl}=\Delta \alpha_J=\Delta \phi,
$$
where $\Delta \cdot$ means the variation of the angle along the loop $\delta$, i.e. a loop of the angular momentum. In conclusion, for a rigid body, the classical rotation number is given by the variation of the angle $\phi$ along $\delta$.

\section{Montgomery phase}\label{appc}
\begin{figure}[h!]
\centering
\includegraphics[scale=0.6,trim=50 60 30 50,clip]{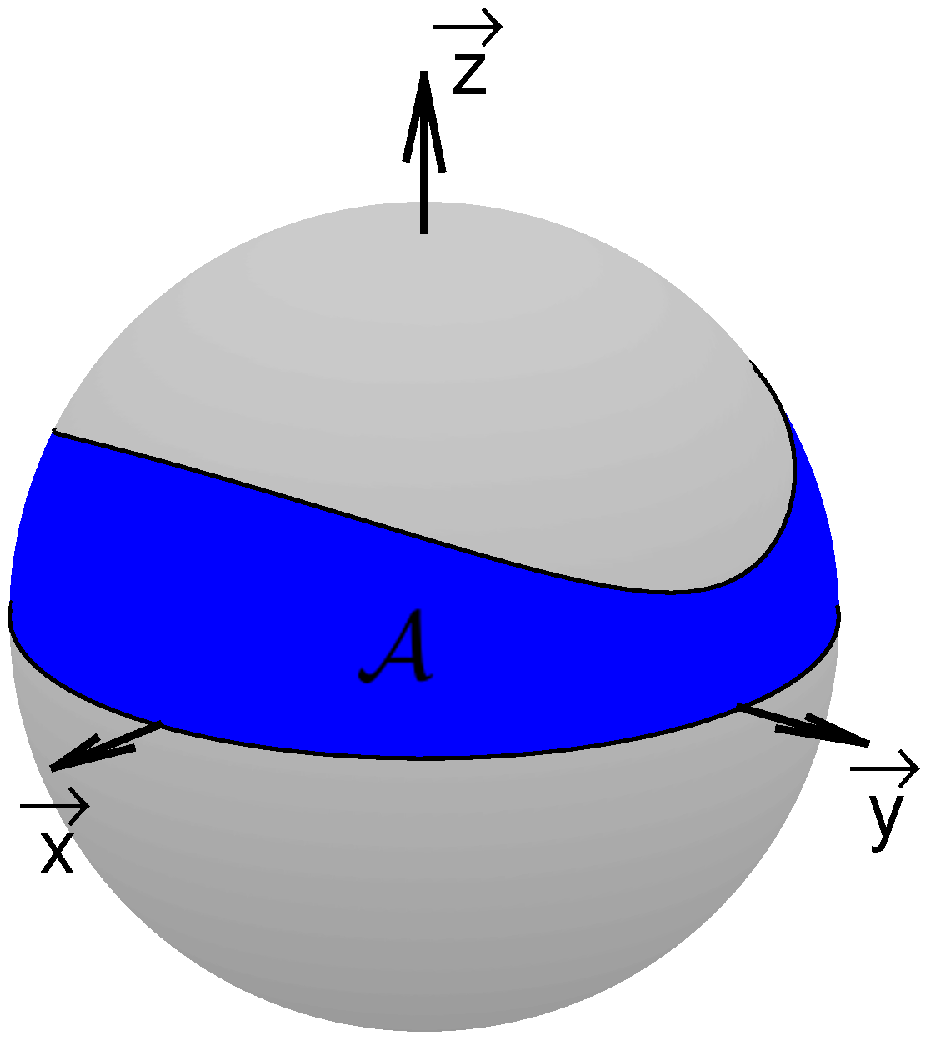}
\includegraphics[scale=0.6,trim=50 60 30 50,clip]{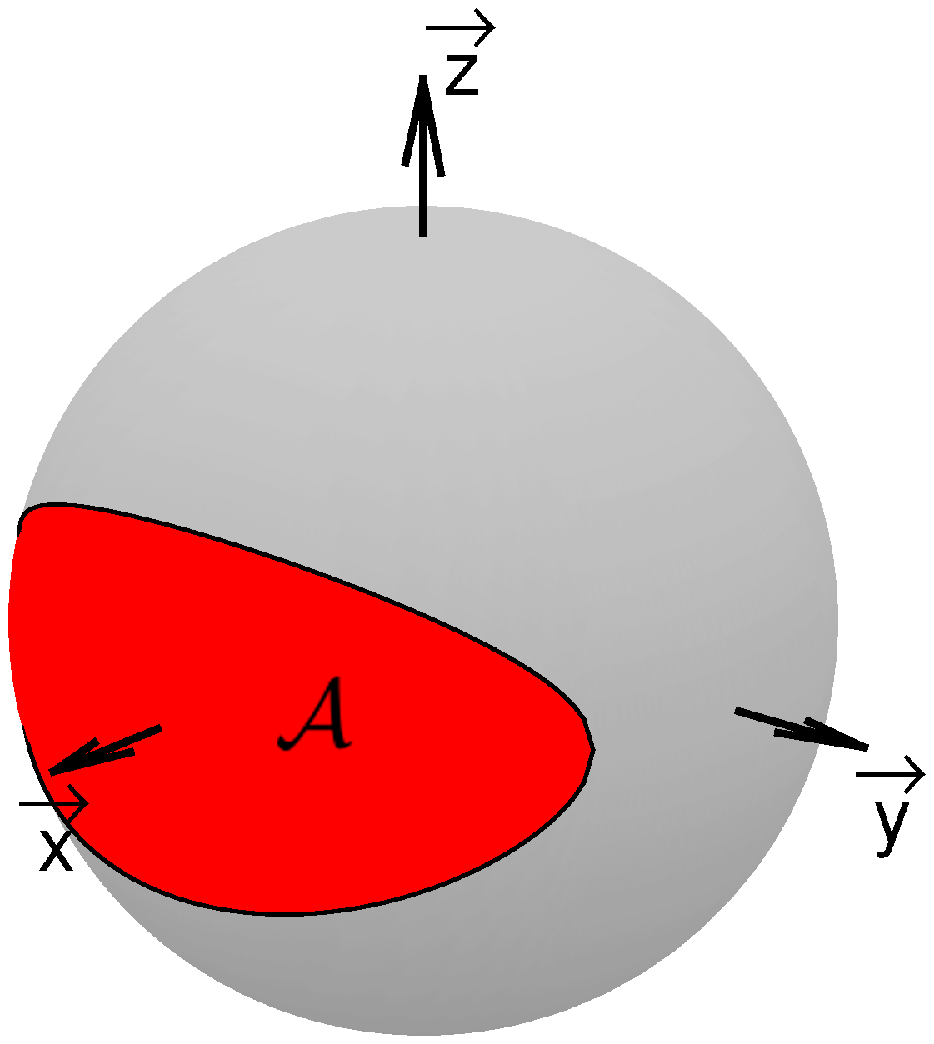}
\caption{(Color online) Plot of the geometric contribution to the Montgomery phase for rotating (top) and oscillating (bottom) trajectories.\label{figSurf}}
\end{figure}

This paragraph is aimed at deriving the Montgomery phase formula~\cite{montgomery,natario}. Using Eq.~\eqref{eqeulerangles} and the expression of the energy $E$ in terms of the angular momentum, we can show that:
\begin{equation}
\dot{\phi}=\frac{2E}{J}-\cos\theta \dot{\psi}.
\end{equation}
The goal is to evaluate the variation of $\phi$ for a period of the angular momentum $J$. For that purpose, we express the angle $\theta$ as a function of $\psi$:
\begin{equation}
\cos^2\theta=\frac{\frac{E}{J^2}-A\cos^2\psi-B\sin^2\psi}{C-A\cos^2\psi-B\sin^2\psi}\label{cos2theta}.
\end{equation}
For a rotating trajectory, we have $\cos\theta>0$ since $\theta\in]0,\pi/2[$ and the angle $\psi$ goes from 0 to $2\pi$. We deduce that:
\begin{equation}
\Delta\phi=\frac{2ET}{J}-\int_0^{2\pi}\cos\theta(\psi) d\psi,
\end{equation}
where $\cos\theta(\psi)$ is given by the square root of Eq.~\eqref{cos2theta}. It is then straightforward to show that:
\begin{equation}\label{eqred}
\Delta\phi=\frac{2ET}{J}-2\int_0^{\pi}\cos\theta(\psi) d\psi.
\end{equation}
The second term of the right-hand side in Eq.~\eqref{eqred} corresponds to the geometric phase, that is the area $\mathcal{A}$ of the surface between the equator of the sphere and the trajectory $\vec{J}(t)$, as shown in Fig.~\ref{figSurf}.
The period $T$ is given by $T=4K(m)/\omega$ with:
\begin{equation}
m=\frac{(B-A)(CJ^2-E)}{(C-B)(E-AJ^2)},~\omega=2\sqrt{(C-B)(E-AJ^2)}.
\end{equation}
In the oscillating case, the term $\cos\theta$ can change of sign. The initial point is defined by $\theta(0)=\pi/2$ which leads to $\psi(0)=\arccos\left(\sqrt{\frac{B-\frac{E}{J^2}}{B-A}}\right)=\psi(T)$. The variation $\Delta\phi$ can be expressed as follows:
\begin{equation}
\Delta\phi=\frac{2ET}{J}-\int_{\psi(0)}^{\pi-\psi(0)}\cos\theta(\psi) d\psi-\int_{\pi-\psi(0)}^{\psi(0)}\cos\theta(\psi)d\psi.
\end{equation}
For oscillating trajectories, the angle $\theta$ takes the value $\pi/2$ when $\psi=\pi-\psi(0)$. For $\psi\in [\psi(0),\pi-\psi(0)]$, $\cos\theta$ can be defined as:
\begin{equation*}
\cos\theta=\sqrt{\frac{\frac{E}{J^2}-A\cos^2\psi-B\sin^2\psi}{C-A\cos^2\psi-B\sin^2\psi}}.
\end{equation*}
For $\psi\in [\pi-\psi(0),\psi(0)]$, we have:
\begin{equation*}
\cos\theta=-\sqrt{\frac{\frac{E}{J^2}-A\cos^2\psi-B\sin^2\psi}{C-A\cos^2\psi-B\sin^2\psi}}.
\end{equation*}
Finally, we obtain:
\begin{equation}
\begin{aligned}
\Delta\phi&=\frac{2ET}{J}-2\int_{\psi(0)}^{\pi-\psi(0)}\cos\theta(\psi) d\psi\\
&=\frac{2ET}{J}-2\int_{\psi(0)}^{\pi-\psi(0)}\sqrt{\frac{\frac{E}{J^2}-A\cos^2\psi-B\sin^2\psi}{C-A\cos^2\psi-B\sin^2\psi}} d\psi.
\end{aligned}
\end{equation}
The geometric phase corresponds to the area displayed in Fig.~\ref{figSurf}. The period $T$ is given by $4K(m)/\omega$ with:
\begin{equation}
m=\frac{(C-B)(E-AJ^2)}{(B-A)(CJ^2-E)},\quad \omega=2\sqrt{(B-A)(CJ^2-E)}.
\end{equation}

We analyze now the evolution of the rotation number close to the separatrix.
The energy $E$ can be expressed as $E=BJ^2(1+\gamma)$, where $|\gamma|\ll 1$.
The different parameters are approximated as follows:
\begin{equation}
\begin{aligned}
& m=1-\frac{\gamma B(C-A)}{(B-A)(C-B)},\\
&\omega=2J\sqrt{(C-B)(B-A)}+\gamma JB\sqrt{\frac{C-B}{C-A}}
\end{aligned}
\end{equation}
for the rotating case, and:
\begin{equation}
\begin{aligned}
& m=1+\frac{\gamma B(C-A)}{(B-A)(C-B)},\\
&\omega=2J\sqrt{(C-B)(B-A)}-\gamma JB\sqrt{\frac{B-A}{C-B}}
\end{aligned}
\end{equation}
for the oscillating one (note that in this latter case we have $\gamma<0$).

Using the asymptotic expansion~\cite{Abrambook}
\begin{equation}
K(1-\varepsilon)=-\frac{1}{2}\ln \varepsilon +\ln 4+ o(\varepsilon),
\end{equation}
For both cases, we arrive at:
\begin{equation}
\begin{split}
& \frac{2ET}{J}\simeq \frac{4B}{\sqrt{(B-A)(C-B)}}\left[-\frac{\ln (|\gamma|)}{2}+\ln 4\right.\\
&\left. -\frac{1}{2}\ln\left(\frac{(C-A)B}{(B-A)(C-B)}\right)\right]+o\left(|\gamma|\ln(\gamma)\right).
\end{split}
\end{equation}
For the geometric contribution $\Delta\phi_g=-\mathcal{A}$, we obtain for $\gamma=0$:
\begin{equation}
\Delta\phi_g = -4\arcsin \left[\sqrt{\frac{B-A}{C-A}}\right].
\end{equation}
In a neighborhood of the separatrix, the classical rotation number can be written as:
\begin{equation}\label{thetaappr}
\Theta_{cl}=\alpha-\beta\ln (|\gamma|),
\end{equation}
where
\begin{eqnarray*}
& & \alpha=-4\arcsin \left[\sqrt{\frac{B-A}{C-A}}\right]\\
& & +\frac{4B\ln 4}{\sqrt{(C-B)(B-A)}}-2B\ln\left[\frac{B(C-A)}{(C-B)(B-A)}\right].
\end{eqnarray*}
and
\begin{equation*}
\beta =\frac{2B}{\sqrt{(C-B)(B-A)}}.
\end{equation*}

\end{document}